\documentstyle[prd,aps,twoside]{revtex}
\begin{document}

\title{Spherically Symmetric Solutions of Gravitational Field Equations in
Kalb-Ramond Background} 
\author{Soumitra SenGupta\footnote{Electronic address: {\em 
soumitra@juphys.ernet.in}}
${}^{(1)}$,  and Saurabh Sur\footnote{Electronic address: {\em 
saurabh@juphys.ernet.in}}${}^{(2)}$}
\address{{\rm $^{1,2}$}Department of Physics, Jadavpur University,
Calcutta 700 032, India}
\maketitle

\begin{abstract}
Static spherically symmetric solution in a background spacetime with 
torsion is derived explicitly. The torsion considered here is identified
with the field strength of a second rank antisymmetric tensor field namely 
the Kalb-Ramond field and the proposed solution therefore has much 
significance in a string inspired gravitational field theory.
\end{abstract}

\vskip .5in

\section{Introduction}

The extension of the geometric principles of general relativity to the 
physics at a microscopic level where matter formation is done by 
elementary particles, characterized by a spin angular momentum 
in addition to the mass, is achieved in Einstein-Cartan theory. In
such a theory the symmetric Christoffel connection is modified with the 
introduction of an antisymmetric tensorial term, known as the spacetime 
torsion, which is presumed to have a direct relationship with spin \cite
{hehl,akr,sab}. Over the years, since Cartan's pioneering works in early 
1920s, there had been numerous interesting questions especially as to how 
the introduction of torsion in spacetime affects the solutions of the 
gravitational field equations under various circumstances. The advent of 
string theory and its emergence as a powerful tool for the quantum 
theoretic unification of all the fundamental forces of nature \cite{gsw} 
has brought a resurgence of interest in spacetimes with torsion and the 
need for getting satisfactory answers to the aforesaid questions is 
enhanced. The field strength corresponding to the second rank antisymmetric 
tensor in the string spectrum is identified as the space time torsion.

In this paper, we aim to study the possible static (i.e., time-invariant) 
spherically symmetric solutions of the gravitational field equations 
in presence of torsion in spacetime. One of the important 
motivations in looking for such solutions is ideally to realise the 
effect of torsion on the electromagnetic waves coming from distant
galactic sources \cite{nr}. The basic underlying theory adopted by us
is that proposed by Majumdar and SenGupta \cite{pmss}, in which a new 
antisymmetric tensor field $B_{\mu \nu}$, identified as the Kalb-Ramond 
(KR) field, is introduced to ensure the U(1) gauge-invariance of the 
electromagnetic theory in a background with torsion. The massless KR 
field is taken to be the possible source of torsion. The strength of this 
KR field $H_{\mu \nu \lambda} = \partial _{[\mu} B_{\nu \lambda ]}$ is 
modified by U(1) Chern-Simons terms originating from the quantum 
consistency of an underlying string theory and the coupling of the KR 
field with torsion is done in a way that the gauge symmetry is preserved 
in the resulting action.

In Section 2, we focus our attention to the action of a purely gravitational 
field theory in a torsioned background. We vary the action, with respect to 
the metric $g_{\mu \nu}$ and as well as to the Kalb-Ramond field $B_{\mu
\nu}$, to obtain two sets of field equations. The variation of the action 
with respect to $g_{\mu \nu}$ merely gives a modified version of the old 
Einstein's equations. The modification comes in a way that the coupling of 
torsion with gravity is reflected in the spin-density tensor, just as matter 
coupling to gravity is reflected in the energy-momentum tensor in the old 
equations. In Section 3, we seek the possible  spherically symmetric 
solutions of the field equations in  vacuum and show that it is not
necessarily static, unlike what we get in  Schwarzschild solution in
absence of torsion. 
However in section 4, we construct an explicit static spherically
symmetric solution with an appropriate Kalb-Ramond background. 

\vskip .5in

\section{Field Equations for gauge-invariant 
	Einstein-Cartan-Kalb-Ramond Coupling}

\vskip .2in

The action is taken to be of the form \cite{pmss}:

\begin{equation}
  S = \int~ d^{4}x \sqrt{-g}\left[\frac{\tilde{R} (g,T)}{\kappa} -
	\frac{1}{2} H_{\mu \nu \lambda} H^{\mu \nu \lambda} +
	\frac{1}{\sqrt{\kappa}} T^{\mu \nu \lambda} H_{\mu \nu \lambda}\right]
\end{equation}

\noindent
where $\tilde{R}(g,T)$ is the scalar curvature for the Einstein-Cartan 
spacetime where the connections contain torsion $T^{\alpha}_{\mu \nu}$: 

\begin{equation}
\tilde{\Gamma}^{\alpha}_{\mu \nu} = \Gamma^{\alpha}_{\mu \nu} -
T^{\alpha}_{\mu \nu}~,
\end{equation}

\noindent
and $\kappa = 16 \pi G$ is the coupling constant. We choose $T^{\alpha \mu 
\nu}$ to be antisymmetric in its all three indices.

Direct calculation shows that the curvature scalar $\tilde{R}(g,T)$ 
of the Einstein-Cartan spacetime is related to that of purely 
Riemannian spacetime by

\begin{equation}
\tilde{R}(g,T) = R(g) + T_{\mu \nu \lambda} T^{\mu \nu \lambda}.
\end{equation}

Moreover, the torsion tensor $T_{\mu \nu \lambda}$, being an 
auxiliary field in Eq.(1), obeys the constraint equation

\begin{equation}
T_{\mu \nu \lambda} = - \frac{\sqrt{\kappa}}{2} H_{\mu \nu \lambda}
\end{equation}

\noindent
which implies that the augmented KR field strength 3-tensor plays
the role of the spin angular momentum density,  considered to be the 
source of torsion \cite{hehl}.

Substituting Eq.(4) in the action (1) and varying the latter with 
respect to $g_{\mu \nu}$ and $B_{\mu \nu}$ respectively, we obtain
the field equations

\begin{equation}
G_{\mu \nu} \equiv R_{\mu \nu} - \frac{1}{2} g_{\mu \nu} R = \kappa
\tau_{\mu \nu} 
\end{equation}

\noindent
and

\begin{equation}
D_{\mu} H^{\mu \nu \lambda} \equiv \frac{1}{\sqrt{- g}} \partial _{\mu}
(\sqrt{- g} H^{\mu \nu \lambda}) = 0 
\end{equation}

\noindent
where $R_{\mu \nu}$ and $G_{\mu \nu}$ are respectively the Ricci tensor
and the Einstein tensor of Riemannian geometry; and $\tau_{\mu \nu}$ is
a symmetric 2-tensor, having direct analogy with the energy-momentum
tensor, and is a clear manifestation of the spin-density tensor in the
field equations for the Einstein-Cartan spacetime. The expression for
$\tau_{\mu \nu}$ is of the form:
  
\begin{equation}
\tau_{\mu \nu} = \frac{1}{\sqrt{- g}} \left [ \frac{ \partial (\sqrt{- g}
{\cal L} _{KR}) }{ \partial g^{\mu \nu} } - \partial _{\alpha} 
\left (\frac{ \partial (\sqrt{- g} {\cal L} _{KR}) }
{ \partial ( \partial _{\alpha} g^{\mu \nu} ) } \right ) \right ]
\end{equation}

\noindent
where ${\cal L}_{KR}$ is the Lagrangian density due to the KR field,
which can be expressed as

\begin{equation}
{\cal L}_{KR} = \frac{3}{4} H_{\alpha \beta \gamma} H^{\alpha \beta \gamma}.
\end{equation}

Substituting Eq.(8) in Eq.(7), we finally obtain

\begin{equation}
\tau_{\mu \nu} = \frac{3}{4} \left( 3 g_{\nu \rho} H_{\alpha \beta \mu}
H^{\alpha \beta \rho} - \frac{1}{2} g_{\mu \nu} H_{\alpha \beta \gamma}
H^{\alpha \beta \gamma} \right).
\end{equation}

Now, by virtue of the symmetric nature of $G_{\mu \nu}$ and $\tau_{\mu \nu}$,
Eq.(5) can have, in general, ten component equations. Also, the totally 
antisymmetric property of $H_{\mu \nu \lambda}$ implies that it has four
independent components, viz., $H_{012}, H_{013}, H_{023}$ and $H_{123}$,
which we denote, for simplicity, as $h_{1}, h_{2}, h_{3}$ and $h_{4}$ 
respectively; the corresponding contravariant components are denoted by
$h^{1}, h^{2}, h^{3}$ and $h^{4}$ respectively. Eq.(6) is a set of six 
independent equations

\begin{eqnarray}
D_{2} h^{1} + D_{3} h^{2} &=& 0\\
D_{1} h^{1} - D_{3} h^{3} &=& 0\\
D_{1} h^{2} + D_{2} h^{3} &=& 0\\
D_{0} h^{1} + D_{3} h^{4} &=& 0\\
D_{0} h^{3} + D_{1} h^{4} &=& 0\\
D_{0} h^{2} - D_{2} h^{4} &=& 0.
\end{eqnarray}

\vskip .5in

\section{Static Spherically Symmetric Solutions}

\vskip .2in

The line element is taken in its most general spherical symmetric form:

\begin{equation}
ds^{2} = e^{\nu (r,t)} dt^{2} - e^{\lambda (r,t)} dr^{2} - r^{2} (d \theta^{2}
+ \sin ^{2} \theta d \phi^{2})
\end{equation}

\noindent
so that the metric tensor is

\begin{equation}
g_{\mu \nu} = diag( e^{\nu}, - e^{\lambda}, - r^{2}, - r^{2} \sin^{2} \theta ).
\end{equation}

The first set of field equations (5) take the form:

\begin{eqnarray}
e^{- \lambda}\left (\frac{1}{r^{2}} - \frac{\lambda'}{r} \right) 
- \frac{1}{r^{2}} &=& 
\bar{\kappa} (h_{1}h^{1} + h_{2}h^{2} + h_{3}h^{3} - h_{4}h^{4}) \\
e^{- \lambda}\left (\frac{1}{r^{2}} + \frac{\nu'}{r}\right ) 
- \frac{1}{r^{2}} &=& 
\bar{\kappa} (h_{1}h^{1} + h_{2}h^{2} - h_{3}h^{3} + h_{4}h^{4}) \\ 
e^{- \lambda}\left (\nu'' + \frac{\nu'^{2}}{2} - \frac{\nu' 
\lambda'}{2} + \frac{\nu' - \lambda'}{r} \right ) 
- e^{- \nu}\left (\ddot{\lambda} + \frac{\dot
{\lambda}^{2}}{2} - \frac{\dot{\lambda} \dot{\nu}}{2}\right ) &=& 
2 \bar{\kappa} (h_{1}h^{1} - h_{2}h^{2} + h_{3}h^{3} + h_{4}h^{4}) \\
e^{- \lambda}\left (\nu'' + \frac{\nu'^{2}}{2} - \frac{\nu' 
\lambda'}{2} + \frac{\nu' - \lambda'}{r}\right ) 
- e^{- \nu}\left (\ddot{\lambda} + \frac{\dot
{\lambda}^{2}}{2} - \frac{\dot{\lambda} \dot{\nu}}{2}\right ) &=& 
2 \bar{\kappa} (- h_{1}h^{1} + h_{2}h^{2} + h_{3}h^{3} + h_{4}h^{4}) \\
\frac{e^{- \lambda} \dot{\lambda}}{r} &=& 2 \bar{\kappa} h_{3}h^{4} \\
h_{2}h^{4} = h_{1}h^{4} = h_{2}h^{3} &=& h_{1}h^{2} = h_{3}h^{1} = 0
\end{eqnarray}

\noindent
and the other set (10) - (15) reduce to:

\begin{eqnarray}
h^{1},_{2} + \cot \theta ~h^{1} &=& - h^{2},_{3} \\
h^{1},_{1} + \left (\frac{\nu' + \lambda'}{2} + \frac{2}{r} \right ) 
h^{1} &=& h^{3},_{3} \\
h^{2},_{1} + \left (\frac{\nu' + \lambda'}{2} + \frac{2}{r} \right ) 
h^{2} &=& - h^{3},_{2}
- \cot \theta ~h^{3} \\
h^{1},_{0} + \left (\frac{\dot{\nu} + \dot{\lambda}}{2} \right ) 
h^{1} &=& - h^{4},_{3} \\
h^{3},_{0} + \left (\frac{\dot{\nu} + \dot{\lambda}}{2} \right ) 
h^{3} &=& - h^{4},_{1} - \left (\frac{\nu' + \lambda'}{2} 
+ \frac{2}{r} \right ) h^{4} \\
h^{2},_{0} + \left (\frac{\dot{\nu} + \dot{\lambda}}{2} \right ) 
h^{2} &=& h^{4},_{2} + \cot \theta ~h^{4}.
\end{eqnarray}

Here dot and prime respectively stand for the differentiations with respect
to $t$ and $r$; and $ \bar{\kappa} = \frac{9 \kappa}{4}$.

The left hand sides of Eqs.(20) and (21) are identical, which implies
that both $h_{1}$ and $h_{2}$ must vanish, and this will satisfy Eq.(23) as
well. It then follows from Eqs.(25) and (27) that

\begin{equation}
h^{3},_{3} = h^{4},_{3} = 0
\end{equation}

\noindent
which means that $h^{3}$ and $h^{4}$ are both independent of the coordinate
$\phi$. But from Eq.(22) we find that $\dot{\lambda}$ is, in general, 
non-zero, i.e., $\lambda$ is time-dependent. 
Thus we arrive at an important result:  the spherically symmetric 
gravitational field in vacuum, in a spacetime with torsion, is not 
necessarily static, unlike the case in the Schwarzschild solution of 
Einstein's field equations. As we are primarily interested in static
vacuum solutions in torsioned background, we set $\dot{\lambda} = \dot
{\nu} = 0$, which means that both $\lambda$ and $\nu$ are time-invariant 
and Eq.(22) suggests that either $h_{3}$ or $h_{4}$ or both 
should be equal to zero. But if both $h_{3}$ and $h_{4}$ vanish, i.e., 
all the torsion components are zero, we get back the Schwarzschild metric; 
and since we are interested in solutions involving torsion, we are bound 
to make the following two choices:

\vskip .2in

\subsection{Choice I : $h_{3} \neq 0, h_{4} = 0$}

In this case the field equations take the form:

\begin{eqnarray}
e^{- \lambda}\left (\frac{1}{r^{2}} - \frac{\lambda'}{r}\right ) 
- \frac{1}{r^{2}} &=& \bar{\kappa} h_{3}h^{3} \\
e^{- \lambda}\left (\frac{1}{r^{2}} + \frac{\nu'}{r}\right ) 
- \frac{1}{r^{2}} &=& - \bar{\kappa} h_{3}h^{3} \\
e^{- \lambda} \left (\nu'' + \frac{\nu'^{2}}{2} - \frac{\nu' 
\lambda'}{2} + \frac{\nu' - \lambda'}{r} \right ) &=& 
2 \bar{\kappa} h_{3}h^{3} \\
h^{3},_{2} + \cot \theta ~h^{3} &=& 0 \\
h^{3},_{0} = h^{3},_{3} &=& 0.
\end{eqnarray}

The left hand sides of Eqs.(31) - (33) are all functions of $r$ only,
therefore, their right hand sides consisting of $h_{3}h^{3}$ must also be
functions of r only. It follows from Eq.(35) that $h_{3}$ (and hence $h^{3}$)
is not only independent of $\phi$, but also independent of $t$.

Now,

\begin{eqnarray}
h_{3} \equiv H_{023} &=& g_{00} ~g_{22} ~g_{33} ~H^{023} \nonumber\\
&=&   e^{\nu} r^{4} \sin^{2} \theta ~h^{3}.
\end{eqnarray}

Hence,

\begin{equation}
h_{3}h^{3} = e^{\nu} r^{4} \sin^{2} \theta ~(h^{3})^{2}.
\end{equation}
 
On integration, Eq.(34) yields

\begin{equation}
h^{3} (r,\theta) = \frac{h^{3}(r)}{\sin \theta}.
\end{equation}

Substituting Eq.(38) in Eq.(37), we get

\begin{equation}
h_{3}h^{3} = e^{\nu} r^{4} [h^{3}(r)]^{2}
\end{equation}

\noindent
which  indicates that $h_{3}h^{3}$ is indeed a function of $r$ only.

We denote

\begin{equation}
h_{3}h^{3} \equiv [h(r)]^{2}
\end{equation}

\noindent
and write the field equations in a more convenient way so that they can be
solved easily:

\begin{eqnarray}
\frac{d}{dr} ~(r e^{- \lambda}) &=& 1 + \bar{\kappa} r^{2} h^{2} \\
(r e^{- \lambda}) \{ \nu' + \frac{d}{dr} [\ln(r^{2} e^{- \lambda})] \}
&=& 2 \\
e^{- \lambda}\left (\nu'' + \frac{\nu'^{2}}{2} - \frac{\nu' \lambda'}{2} +
\frac{\nu' - \lambda'}{r}\right ) &=& 2 \bar{\kappa} r^{2} h^{2}
\end{eqnarray}

\noindent
where the second equation, i.e., Eq.(42), is obtained simply by adding 
Eqs.(31) and (32). Eqs.(41) and (42) can be solved simultaneously to obtain

\begin{equation}
e^{- \lambda} = 1 + \frac{c_{1}}{r} + \frac{\tau (r)}{r}
\end{equation}

\noindent
and
 
\begin{equation}
e^{\nu} = \frac{c_{2}}{r(r + \tau (r) + c_{1})} \exp \left[ 
\int ~ \frac{2 dr}{r + \tau (r) + c_{1}} \right]
\end{equation}

\noindent
where $c_{1}$ and $c_{2}$ are the constants of the integrations and
 
\begin{equation}
\tau (r) = ~\bar{\kappa} ~\int ~r^{2} h^{2} (r) dr. 
\end{equation}

For vanishing torsion (i.e., $\tau(r) = 0$), these solutions 
reduce exactly to the Schwarzschild solution, viz., $~e^{\nu}~ 
= ~e^{- \lambda} ~=~ 1 - \frac{r_{s}}{r}$, provided $c_{2} = 1$ 
and $c_{1} = - r_{s}$, where $r_{s} = 2 G M$ is the Schwarzschild 
radius. However, when torsion is non-zero, the acceptability of these
solutions 
rests on two factors: ~firstly, despite satisfying Eqs.(41) and (42), 
they must satisfy Eq.(43) as well, which is a separate field equation 
involving $\nu$ and $\lambda$; and secondly, they must attain the 
asymptotic forms $~e^{\nu} = e^{\lambda} = 1~$ in the limit $~r 
\rightarrow \infty$. Both these conditions may be fulfilled only with 
some suitable form of $\tau(r)$. We explore such a possibility in 
Section 4. 

\vskip .2in

\subsection{Choice II : $h_{4} \neq 0, h_{3} = 0$}

The field equations for this choice are

\begin{eqnarray}
e^{- \lambda}\left (\frac{1}{r^{2}} - \frac{\lambda'}{r}\right ) 
- \frac{1}{r^{2}} &=& - \bar{\kappa} h_{4}h^{4} \\
e^{- \lambda}\left (\frac{1}{r^{2}} + \frac{\nu'}{r}\right ) 
- \frac{1}{r^{2}} &=& \bar{\kappa} h_{4}h^{4} \\
e^{- \lambda}\left (\nu'' + \frac{\nu'^{2}}{2} - 
\frac{\nu' \lambda'}{2} + \frac{\nu' - \lambda'}{r}\right ) 
&=& 2 \bar{\kappa} h_{4}h^{4} \\
h^{4},_{1} +\left (\frac{\nu' + \lambda'}{2} + \frac{2}{r}\right ) 
h^{4} &=& 0 \\
h^{4},_{2} + \cot \theta ~h^{4} &=& 0 \\
h^{4},_{3} &=& 0.
\end{eqnarray}

Here, again we find from Eqs.(47) - (49) that $h_{4}h^{4}$ must have to be
a function of $r$ only, i.e., independent of $t, \theta$ and $\phi$. Eq.(52)
shows that $h_{4}$, and hence $h^{4}$, is independent of $\phi$, but there
is no equation from which it is evident that $h_{4}h^{4}$ is independent of
$t$. Meanwhile, we have

\begin{eqnarray}
h_{4} \equiv H_{123}  &=& g_{11} ~g_{22} ~g_{33} ~H^{123} \nonumber \\
&=& - e^{\lambda} r^{4} \sin^{2} \theta ~h^{4}
\end{eqnarray}

\noindent
which means 

\begin{equation}
h_{4}h^{4} = - e^{\lambda} r^{4} \sin^{2} \theta ~(h^{4})^{2}.
\end{equation}
 
Integrating Eq.(51), we get 

\begin{equation}
h^{4} (r, \theta, t) = \frac{h^{4}(r,t)}{\sin \theta}.
\end{equation}

This on substitution into Eq.(54), yields 

\begin{equation}
h_{4}h^{4} = - e^{\lambda} r^{4} [h^{4}(r,t)]^{2}
\end{equation}

\noindent
which is independent of $\theta$. We assume that $h_{4}h^{4}$ is 
independent of $t$ as well, so that a static spherically symmetric 
solution can be obtained:

\begin{equation}
h_{4}h^{4} \equiv [\bar{h}(r)]^{2}.
\end{equation}

The $r$-dependence of $h^{4}$ could be found directly from Eq.(50): 

\begin{equation}
h^{4} (r,t,\theta) = \frac{h^{4} (t,\theta)}{r^{2}} e^{- \frac{\nu +
\lambda}{2}} 
\end{equation}

\noindent
so that

\begin{equation}
h_{4}h^{4} = - e^{- \nu} [h^{4} (r,t)]^{2} = - \alpha e^{- \nu}
\end{equation}

\noindent
where we take $[h^{4} (t)]^{2} = \alpha$ (a constant), i.e., $h^{4}$
is kept time-invariant so as that $h_{4}h^{4}$ remains independent of
time.

Thus

\begin{equation}
\bar{h} (r) = i ~\sqrt{\alpha} ~e^{-\nu /2}
\end{equation}

But this is quite unrealistic and not acceptable since $e^{- \nu}$, identified
as $g^{00}$, must approach unity in the asymptotic limit $~r \rightarrow
\infty~$, leading to a finite non-vanishing value of torsion in that limit
as is evident from Eq.(59). Moreover, Eq.(47) gives the result 

\begin{equation}
g^{00} \equiv e^{-\nu} = \frac{ \frac{d}{dr}(r e^{-\lambda}) - 1}{\bar{\kappa}
\alpha r^{2}},
\end{equation}

\noindent
wherefrom it is clear that in order to make $g^{00}$ approach unity 
asymptotically as $~r \rightarrow \infty~$ we must choose 

\begin{equation}
\frac{d}{dr}(r e^{-\lambda}) = 1 + \bar{\kappa} \alpha r^{2} \gamma(r) + 
\delta(r),
\end{equation}

\noindent
and impose the conditions on the arbitrary functions $\gamma(r)$ and $\delta(r)$
that they must approach $1$ and $0$ respectively in the $~r \rightarrow \infty~$
limit. But this is simply not possible since this implies that $e^{-\lambda}$, 
i.e., $-g^{11}$ tends to infinity (instead of approaching unity) as $~r 
\rightarrow \infty~$.

Hence, we disregard the choice II and infer that the choice I leads to the only 
possibility of getting a generalized spherically symmetric solution of the 
gravitational field equations in a curved spacetime with a Kalb-Ramond background.

\vskip .5in

\section{ Exact static spherically symmetric solutions for the metric and
the Kalb-Ramond background}

\vskip .2in

A close insight to the field equations (41) - (43) reveals 
that the solutions (44) and (45), which were obtained by solving Eqs.
(41) and (42) only, would satisfy Eq.(43) as well, only when the function 
$\tau (r)$, which depends on the KR field strength $h(r)$, satisfies the 
equation 

\begin{equation}
\tau'' + \frac{\tau'}{r} = \frac{\tau' (\tau' - 1)}{r + c_{1} + \tau}
\end{equation}

\noindent
which, on first integration, yields

\begin{equation}
\tau' = \frac{\alpha}{r} (r + c_{1} + \tau) \exp \left [- \int ~\frac{2 ~dr}
{r + c_{1} + \tau} \right ]
\end{equation}

\noindent
where $\alpha$ is an integration constant.

The above equation has an exact solution

\begin{equation}
\tau(r) = \frac{\alpha}{r^{2}}
\end{equation}

\noindent
for $c_{1} = 0$, whence we obtain the solution consistent with all the 
three field equations (41) - (43)~:
 
\begin{eqnarray}
e^{- \lambda} &=& 1 - \frac{\alpha}{r^{2}} \\
e^{\nu} &=& 1.
\end{eqnarray}

Here we note that the constant $c_{2}$ in eq.(45) is taken to be equal to $1$ 
in order that $e^{\nu}$ remains unity when $~r \rightarrow \infty$. Moreover,
the form of $\tau(r)$ in Eq.(65) suggests that the KR field strength has an
inverse square dependence on $r$:

\begin{equation}
h(r) = \sqrt{\frac{\alpha}{\bar{\kappa}}} \frac{1}{r^{2}}.
\end{equation}

But depending on the sign of $\alpha$, we have the following two cases:

{\bf Case I : \/}
For real and positive $\alpha$ the KR field $h$ is real and we cannot
find any event horizon for the metric under consideration. This implies 
that the singularity at $r = 0$ is, actually, a naked singularity, not being
shielded by an event horizon as in the ideal Schwarzschild case. In fact,
there exists another singularity at $r = \sqrt{\alpha}$, whose origin
is not quite clear, at least physically, but it is certain that its presence 
cannot lead to an event horizon.

{\bf Case II : \/}
For real and negative $\alpha$ the KR field $h$ is imaginary, 
but again we find the absence of an event horizon leads to a naked 
singularity at $r = 0$. Moreover, we note that, unlike the case I, 
we do not have any other singularity in this case.

\vskip .5in

\section{Conclusions}

\vskip .2in

We, therefore, have shown that in a spherically symmetric Kalb-Ramond
background one may obtain a static spherically symmetric solution for
the gravitational field equations. The KR field, modified suitably by
Chern-Simons terms for gauge-invariance, may couple with electromagnetic
field to produce non-trivial effects like optical activity for the 
radiowaves from the distant galaxies \cite{skpm,pdpj}. In the 
references cited, the calculations have been done for a flat universe 
and more generally for matter and radiation dominated universe. However, 
our work will now allow to estimate such effects in static spherically 
symmetric background with KR field.

As the augmented KR background can be obtained in the scenario of string
theory in the form of a second-rank antisymmetric tensor field \cite{gsw}
from a completely different angle like anomaly cancellation, our work 
therefore proposes a static spherically symmetric solution for string 
inspired gravity. In string theory the field strength corresponding to the
second rank antisymmetric tensor field ( which is identified with
spacetime torsion ) is related to
the axion through a duality. Our work therefore essentially exhibits the
existence of static spherically symmetric solution in an axion background.
This solution now opens new possibilities of exploring
various aspects of the gravitational solution including black holes. Work 
in this direction is in progress. Furthermore, the nature of the nonstatic 
solutions and their implications must also be investigated to have a 
complete understanding of the spherically symmetric solution in a 
torsioned background. We also propose to extend this work in a more
generalized scenario where the Kalb-Ramond background admits of a parity
violating extension \cite{bmss}. We expect to have some interesting 
solutions in such a case which may offer an explanation to the parity
violating phenomenon in such a spacetime.

It must also be pointed out here that the horizon-free solution 
obtained for the inverse square nature  of the KR field is completely
compatible with the "no hair" conjecture \cite{bm}. That is to say this 
form of KR field exhibits an important feature in having no "hair" 
whatsoever.

\vskip .5in

\section{Acknowledgement}

\vskip .2in
We acknowledge illuminating discussions with N. Banerjee.
This work is supported by Project grant no. 98/37/16/BRNS cell/676 from The 
Board of Research in Nuclear Sciences, Department of Atomic Energy, Government
of India, and the Council of Scientific and Industrial Research, 
Government of India.


\vskip .5in

\begin{references}
\bibitem{hehl} Hehl F, von der Heyde P, Kerlick G and Nester J,  
		Rev. Mod. Phys.{\bf 48} 393 (1976).
\bibitem{akr} Raychaudhuri A K, Theoretical Cosmology, 1979 (Clarendon 
		Press, Oxford). 
\bibitem{sab} de Sabbata V and Gasperini M, Introduction to Gravitation, 1985
		(World Scientific, Singapore).
\bibitem{gsw} Green M, Schwarz J and Witten E, Superstring Theory v.2, 1985 
		(Cambridge: Cambridge Univ. Press). 
\bibitem{nr} Nodland B and Ralston J P, Phys. Rev. Lett. {\bf 78},
		3043 (1997) ; {\em ibid} Phys. Rev. Lett. {\bf 79}, 1958 (1997);
		astro-ph/9708114;astro-ph/9706126.
\bibitem{pmss} Majumdar P and SenGupta S, Class. Quan. Grav.
		{\bf 16} L89 (1999).  
\bibitem{skpm} Kar S, Majumdar P, SenGupta S and Sinha A, 2000, {\it Preprint}
		gr-qc/0006097. 
\bibitem{pdpj} Das P, Jain P and Mukherji S, 2000, {\it Preprint} hep-ph/
		0011279.
\bibitem{bmss} Mukhopadhyaya B and SenGupta S, 1998, Phys. Lett. B 
		{\bf 458}, 8 - 12 (1999)
\bibitem{bm} Mayo A E and Bekenstein J D, Phys. Rev. D {\bf 54}, 5509 (1996)
		[and the references therein].
\end{references}
\end{document}